# Perpendicular Magnetization and Generic Realization of the Ising Model in Artificial Spin Ice

Sheng Zhang,[1] Jie Li,[1] Ian Gilbert,[1] Jason Bartell,[1] Michael J. Erickson,[2] Yu Pan,[1] Paul E. Lammert,[1] Cristiano Nisoli,[3] K. K. Kohli,[1] Rajiv Misra,[1] Vincent H. Crespi,[1] Nitin Samarth,[1] C. Leighton,[2] and Peter Schiffer[1*]

[1]*Department of Physics and Materials Research Institute, Pennsylvania State University, University Park, PA 16802, USA*

[2]*Department of Chemical Engineering and Materials Science, University of Minnesota, Minneapolis, MN 55455, USA*

[3]*Theoretical Division and Center for Nonlinear Studies, Los Alamos National Laboratory, Los Alamos, NM 87545, USA*

We have studied frustrated kagome arrays and unfrustrated honeycomb arrays of magnetostatically-interacting single-domain ferromagnetic islands with magnetization normal to the plane. The measured pairwise spin correlations of both lattices can be reproduced by models based solely on nearest-neighbor correlations. The kagome array has qualitatively different magnetostatics but identical lattice topology to previously-studied 'artificial spin ice' systems composed of in-plane moments. The two systems show striking similarities in the development of moment pair correlations, demonstrating a universality in artificial spin ice behavior independent of specific realization in a particular material system.

Frustration in magnetic systems has long been known to generate novel phenomena[1, 2], ranging from spin liquids in which atomic moments fluctuate as the temperature approaches absolute zero[3] to spin ices with monopole-like excitations[4-7]. A new manifestation of magnetic frustration has been examined recently in artificial frustrated magnets, systems wherein the magnetic moments of lithographically patterned ferromagnetic films are arranged so that their magnetostatic interactions are frustrated[8]. These systems have been most closely studied in the context of ice-like geometries (i.e., 'artificial spin ice') and have opened a new avenue in the study of frustration, since the interactions are controllable and the local moment arrangements are directly observable[9-19]. We have studied a new form of artificial frustrated magnet, consisting of magnetostatically-interacting single-domain ferromagnetic islands with moments oriented perpendicular to the plane, rather than in-plane as in all previous studies. In particular, we examine a kagome geometry with qualitatively different magnetostatics but identical lattice topology to previously-studied 'artificial spin ice' systems with in-plane moments, and the two systems show striking similarities in the development of moment pair correlations. Furthermore, we demonstrate that *both* systems closely follow expectations for a nearest-neighbor Ising model, indicating a universality in artificial spin ice behavior independent of specific realization in a particular material system.

Our samples were fabricated from multilayer metallic thin films with structure Ti(20Å) / Pt(100Å) / [Co(3Å)/Pt(10Å)]$_8$ deposited by electron-beam evaporation after electron-beam patterning of a bi-layer resist; a similar method was used previously to produce small hexagonal island clusters[20]. The films were characterized structurally via Grazing Incidence X-ray Reflectivity (GIXR) and Wide-Angle X-ray Diffraction (WAXRD). GIXR revealed the expected oscillations due to the full stack thickness plus a first-order superlattice peak from the [Co/Pt] superlattice. The structure and in-plane texturing are consistent with previous studies of such superlattices; (more details are given in Supplementary information)[21]. These multilayers are known to have sufficient interface-induced perpendicular magnetic anisotropy to induce an out-of-plane easy axis of the magnetization[22, 23], and SQUID magnetometry data confirm that the easy axis of magnetic moment for unpatterned films is out of the plane[21].

As shown in Fig. 1, we patterned our films into honeycomb and kagome lattice arrays of circular islands (of radius 200 nm and moment $\sim 5.9 \times 10^7$ $\mu_B$)[23], with nearest-neighbor separations varying from 500 to 1200 nm. The magnetostatic interactions between all pairs of islands are antiferromagnetic and isotropic within the plane, and they depend only on the island separation. To generate a low-energy magnetostatic state, the samples were subjected to an ac demagnetization protocol similar to that used in previous artificial spin ice studies[8, 9, 14, 15, 24]. Each sample was rotated at 1000 rpm while a magnetic field of 2000 Oe was applied perpendicular to an in-plane rotational axis (aligned with the vertical direction in Fig. 1) and stepped down to zero in 1.6 Oe increments, reversing polarity at each step. After demagnetization, we used magnetic force microscopy (MFM) to map the resulting moment configuration, as shown in Fig. 1. Each island is uniformly black or white, reflecting a single domain with moment perpendicular to the plane of the sample. MFM images were taken at five different locations within each array. The number of islands imaged at each location varies from 150 to 1100 depending on the lattice parameter.

The MFM data reveal interaction-induced correlations among the moments. We define the pair correlation as the empirical average spin product (+1 for anti-parallel moments, -1 for parallel) over all pairs of each geometrically distinct pair type. Error bars are calculated under an assumption of independence, and consequent binomial distribution of the total tabulated sample populations for each lattice/spacing type. In the case of nearest-neighbor pairs in the kagome lattice (as well as the in-plane hexagonal lattice mentioned later), geometrical constraints prompt us to take the elementary triangles (one or three frustrated bonds) as population constituents. But in all other cases, it is simply pairs (aligned or anti-aligned). For the arrays with largest lattice spacing, the correlations nearly vanish, even between nearest neighbors, but they increase substantially for the denser lattices. We focus mainly on the smallest lattice spacing, 500 nm, since it exhibits the strongest interaction effects. Figs. 2a,b show the correlations between different neighbor pairs, arranged in order of the magnetostatic energies (given in Fig. 2d,e) calculated through micromagnetic simulations[21, 25]. In each case the nearest neighbor correlations are largest, which is unsurprising given that this interaction is much stronger than the next-nearest neighbor one.

To better understand the source of the observed correlations, we must consider that the observed moment configurations are outside of thermal equilibrium, and that the demagnetization process by which they reach a low energy state is not observed here. Standard thermodynamics in a Gibbsian framework produces a maximum entropy state subject to certain experimentally observable macroscopic constraints such as volume or pressure. Artificial spin systems are unusual in that microstates are directly observable, and hence a broader selection of possible constraints, some microscopic, are available[12, 26-29]. We take the experimentally observed nearest-neighbor pair correlation as the constraint, and we apply two distinct approaches to its imposition: a quasi-equilibrium Gibbsian model (Model G) and a kinetic zero-temperature quenched model (Model Z), assuming ideal Ising spins in both cases (details in Supplementary information[21]). The Gibbsian model takes the probability of a configuration to be $\exp\Phi(s)$, with a nearest-neighbor interaction:

$$\Phi(s) = K \sum_{\text{NN pairs}} s_i s_j$$

This state can be calculated by standard Monte Carlo methods, with $K$ adjusted to match the experimental nearest-neighbor correlation, in order to produce the maximum entropy state consistent with a given nearest-neighbor correlation[26]. The quenched model is purposefully constructed with limited kinetics and a nonphysical starting point. It starts with a completely random moment configuration and flips randomly selected moments only if doing so lowers the nearest-neighbor interaction energy, continuing until the nearest-neighbor correlation matches the experimental value. As seen in Fig. 2, both models reproduce the experimental data well, with significant deviations only for the furthest neighbors in the honeycomb lattice. (For these further neighbors, Model G overestimates the correlations and hence Model Z performs better. Due to the simplicity of both models, one should be cautious in interpreting this difference. Nevertheless, a Gibbsian description of Model Z [30] does have an effective four-spin interaction which opposes long-range antiferromagnetic ordering. Alternatively, quenched disorder in the island switching fields due to small variations in shape may impede long-range antiferromagnetic ordering. Both models lack quenched disorder, but the non-dynamical Model G may more vulnerable to this deficiency, since Model Z at least begins from a random initial state. One should also keep in mind that the real source of the suppression may be long-range dipolar interactions which are absent in both models.) The substantial agreement between two such disparate models and the experiments suggest that the collective state of the perpendicular moment systems is effectively driven only by the nearest neighbor correlations and the lattice topology, i.e., the nearest-neighbor correlation constraint is a robust single physical measure that characterizes the outcome of the rotational demagnetization.

The kagome and honeycomb lattices have similar geometries (kagome is essentially a honeycomb lattice with overlapping triplets of islands rather than single islands at each vertex), and they do not appear so different from each other on the basis of pairwise correlations plotted in Fig. 2a,b. However, the equilateral triads of the kagome lattice generate local frustration, whereas the honeycomb lattice, with its two equivalent sublattices, is unfrustrated. If nearest neighbor interactions dominate, then the honeycomb lattice has an ordered ground state with antiferromagnetically-aligned sublattices and a simple two-fold spin-flip degeneracy. Indeed, we see domains of ordered moments (i.e., clusters of islands whose moments are locally ordered in the same ground state), as colored in Fig. 3. The typical domain size increases with decreasing lattice spacing, as expected for an interaction effect and can be modeled well by the simulations described above[21]. Such ordering has also been achieved in the initial growth of in-plane square ice[16] and also in a low-symmetry triangular lattice with more complex interactions[9, 14].

In contrast to the honeycomb lattice, our kagome lattice is frustrated and is topologically equivalent to an array of in-plane moments along the *sides* of a hexagonal lattice (compare the neighbor pairing "spider" diagrams of Fig. 2e,f), which is perhaps the most extensively studied of the artificial spin ice systems[10, 11, 13, 15, 17, 26-29, 31]. The mapping between these lattices requires a sign convention for the in-plane moments. The vertices of the hexagonal lattice comprise two sub-lattices; we define a moment as positive if it points towards one of them and negative if it points toward the other. With this sign convention, both lattices have effectively antiferromagnetic nearest-neighbor interactions, and spin correlations for any pair can be consistently compared between the two lattices. Fig. 2c plots these correlations for an in-plane hexagonal lattice previously studied (with a lattice constant of 750 nm) and treated by a similar demagnetization protocol[11], to be compared with the panel 2b just above. The similarity is striking, considering that the two lattices differ qualitatively in the characteristics of the interactions beyond first neighbors: the perpendicular kagome has isotropic, uniformly antiferromagnetic interactions between all pairs while the in-plane hexagonal lattice has mixed effective ferromagnetic and antiferromagnetic interactions that vary with relative island orientation. In addition, these two lattices interact very differently with the external field applied during rotational demagnetization. In the perpendicular lattice every island is aligned identically to the instantaneous direction of the applied field, whereas the in-plane lattice contains three sub-populations of islands with different instantaneous angles to the external field. The quasi-equilibrium Gibbsian (G) and kinetic zero-temperature quenched (Z) models are also able to reproduce these results. (Note that the limited kinetics of the quenched model cannot generate nearest-neighbor correlations approaching 0.33 due to an inability to surmount kinetic barriers against removing residual defects, so the quenched model will fail to describe the most strongly correlated lattices of reference [11]). This close similarity strongly suggests that the physics of the in-plane artificial spin ice system is also dominated by lattice topology and nearest-neighbor interactions.

The striking similarity between the pair correlations of in-plane hexagonal and perpendicular kagome lattices apparent in Fig. 2b,c for strongly interacting lattices naturally motivates an investigation of how the similarity evolves with the strength of the inter-island interactions, which is tunable via the lattice spacing. Fig. 4a plots the nearest neighbor correlations for both systems as a function of nearest neighbor interaction energy across a wide range of lattice spacings. Again, these two lattices display very similar behavior, with correlations abruptly appearing above a threshold interaction strength, increasing at a roughly logarithmic rate, with similar slopes, then saturating at the geometrical maximum correlation of 1/3. (Note that the kagome lattice does not quite reach saturation, but presumably would for a sufficiently dense lattice).

The kinetics of our arrays as the observed state is approached are governed not only by the nearest-neighbor interaction energy, but also by the field-step Zeeman energy M∆H (indicated by marks at the top of Fig. 4a), and a disorder energy scale set by the variations in interactions and individual islands' coercivities due to lithographic and growth inhomogeneities. A physical process governed by three energy scales is unlikely to be well-described by a single-parameter model across its full range of behavior, and therefore one might expect these data to be difficult to model without detailed consideration of the dynamics. Nevertheless, following previous thermodynamic approaches[16, 28, 29] we performed a Monte Carlo simulation for an ideal nearest-neighbor Ising kagome antiferromagnet thermalized at a fixed effective temperature, $T_{eff}$, shown as the solid line in Fig. 4b (details given in Supplementary information[21]). The result successfully reproduces the overall slope of the experimental data (scaled by a constant factor of $T_{eff}$) as the correlation transitions from zero (at high temperatures) to one third (at zero temperature). The parameter $T_{eff}$ is $3.3 \times 10^5$ K and $7.9 \times 10^4$ K for the perpendicular and in-plane systems respectively, values of the same order as the interaction energies. The simulation agreement is not perfect in that the simulation result fails to capture the abrupt onset of correlation for weak interactions, which appears in experiment to be a threshold effect rather than a gentle asymptote to an uncorrelated state. This threshold presumably arises from the demagnetization process. When the field-step Zeeman energy substantially exceeds the nearest-neighbor interaction energy, each island freezes into the random orientation preferred by its individual coercivity; interactions have no ability to control island orientation. Hence the observed correlation falls rapidly to zero when the interaction energy becomes too small. The high-correlation behavior, which also deviates somewhat from the simulation, is likely to be governed by the other ratio: interaction strength versus disorder. Interaction effects must overpower the intrinsic disorder in island coercivity in order to saturate the correlation at one third, and the precise form of that saturation is presumably governed by the distribution of coercivities and interaction energies[18, 19, 22, 23, 32].

The collapse of the experimental data for the two different types of moments, and the agreement with simulation, clearly demonstrate that the physics of artificial spin ice transcends the particular material realization, and even the geometry of the moments. Furthermore, the demonstration of frustrated lattices with moments perpendicular to the plane opens a number of intriguing possibilities for further studies. Perpendicular moments could imprint a frustrated magnetic topology onto the transport properties of thin films underneath the moments, leading to potentially exciting results in systems as diverse as superconductors and 2D electron gases. The perpendicular moment systems also open the possibility of connecting the results from artificial spin ice systems with recent efforts in patterned recording media, and they offer the possibility of applying lessons learned from the physics of frustration to memory or device technologies based on interacting nanomagnets[33-35].

This research has been supported by the U.S. Department of Energy, Office of Basic Energy Sciences, Materials Sciences and Engineering Division under Award # DE-SC0005313, NSF grant DMR-0701582 for support of undergraduate research efforts and lithography has been performed with the support of the National Nanotechnology Infrastructure Network. Work at UMN was supported by the NSF Materials Research Science and Engineering Center under award number DMR-0819885. Theoretical analyses were supported in part by the NSF Materials Research Science and Engineering Center under award number DMR-0820404. We are also very grateful for theoretical input from Claudio Castelnovo and Roderich Moessner.

Figure 1. SEM and MFM images of perpendicular moment nanomagnet arrays in kagome and honeycomb geometries with 600 nm lattice spacing. Each island in MFM shows either black or white, indicating that it consists of a single magnetic domain with moment pointing either up or down.

Figure 2. (a),(b),(c) Correlations as a function of pair order after ac demagnetization for a perpendicular honeycomb lattice at 500 nm spacing, a perpendicular kagome lattice at 500 nm spacing and an in-plane hexagonal lattice at 750 nm spacing (from [11]). Simulated results that constrain the nearest neighbor correlations in a quasi-equilibrium Gibbsian model (Model G: Red circles) and a kinetic zero-temperature quenched model (Model Z: Green triangles) agree well with the experiment. (d),(e),(f) Corresponding pair energies from micromagnetic simulations [25] as a function of pair order, using black points for antiferromagnetic interactions and red points for ferromagnetic interactions. Note that AF/FM interactions are indicated only for the in-plane lattice because the interactions are purely antiferromagnetic in the perpendicular lattice material. The insets label the neighbor pairs for each lattice.

Figure 3. (a,b) Ground state domains in perpendicular honeycomb lattices at 500 nm and 800 nm spacing in the same image scale ($18 \times 18$ μm$^2$). Red and blue areas indicate the two-fold ground state degeneracy.

Figure 4. (a) Nearest neighbor correlations as a function of nearest neighbor interaction energies for all inter-island spacings of the perpendicular kagome lattices and in-plane hexagonal lattices. Marks at the top are the field-step Zeeman energy MΔH. (b) Monte Carlo simulation results for the ideal Ising kagome antiferromagnet and data from the main figure scaled to match. E is the nearest neighbor interaction energy, and $k_B$ is Boltzmann constant.

Figure 1

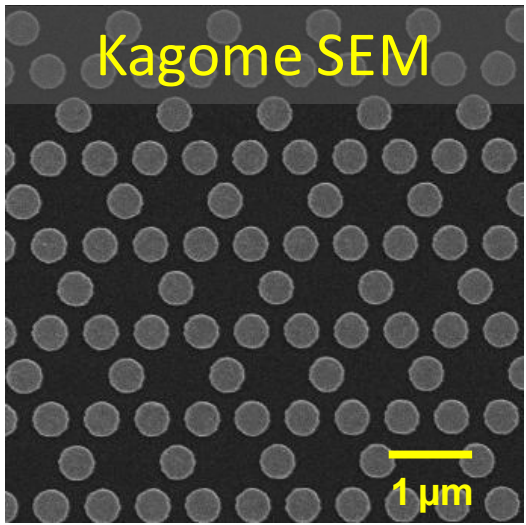 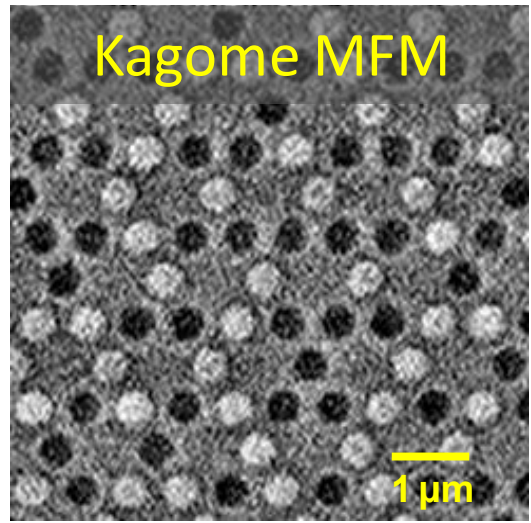

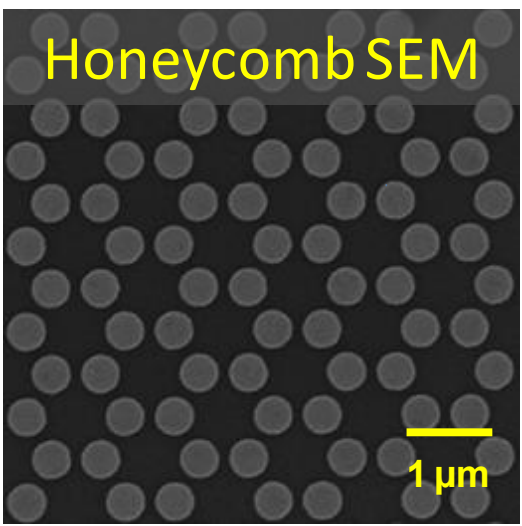 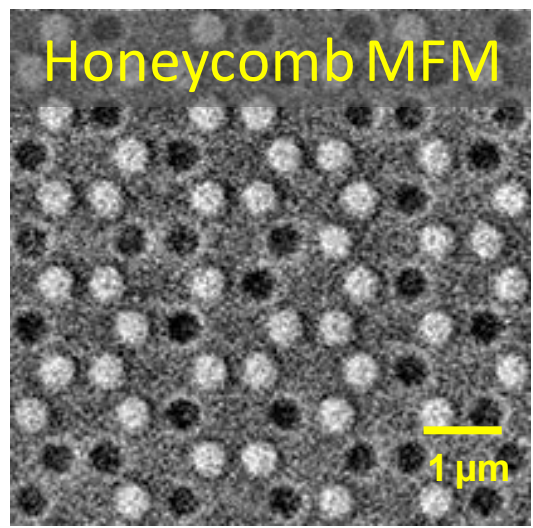

Figure 2

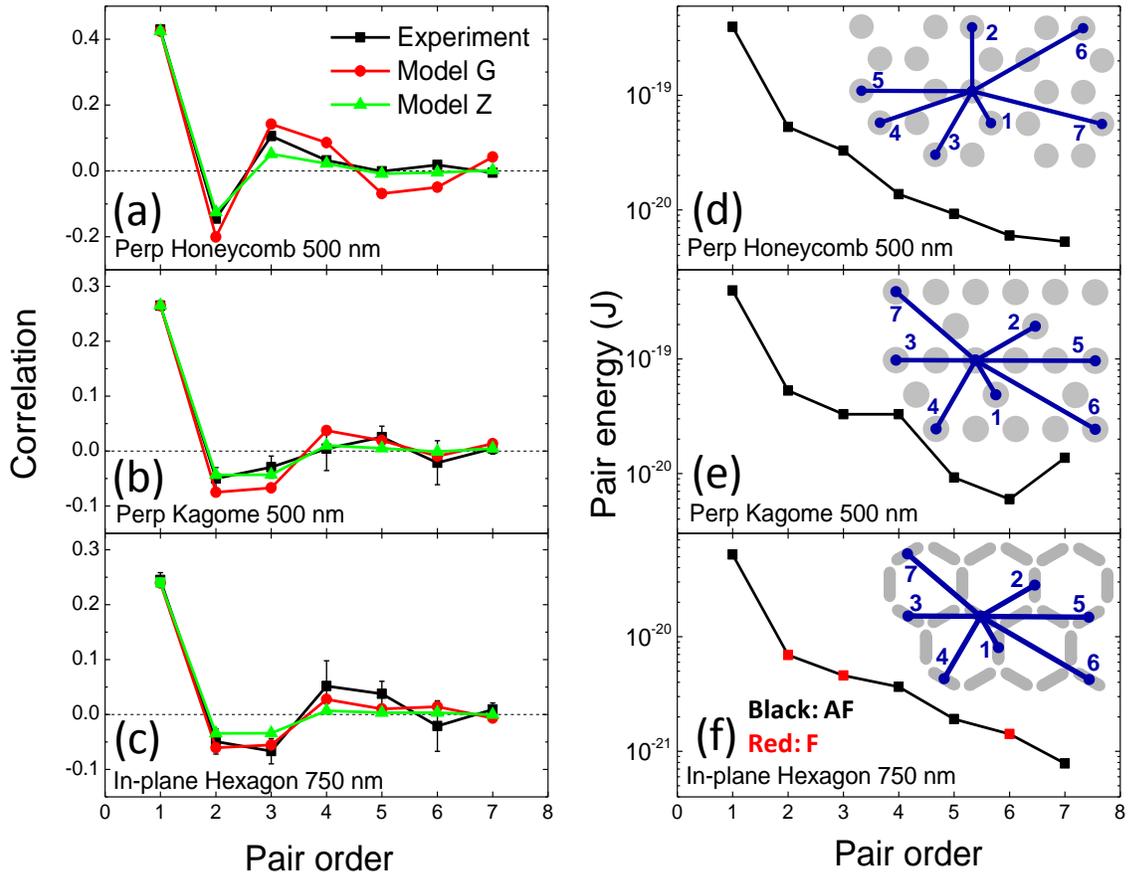

Figure 3

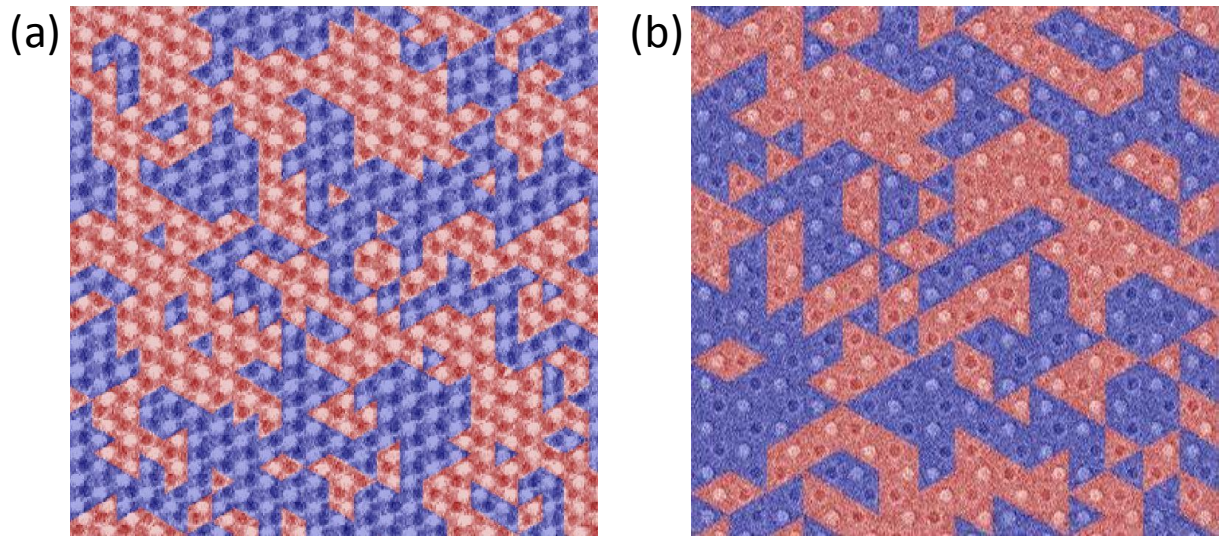

Figure 4

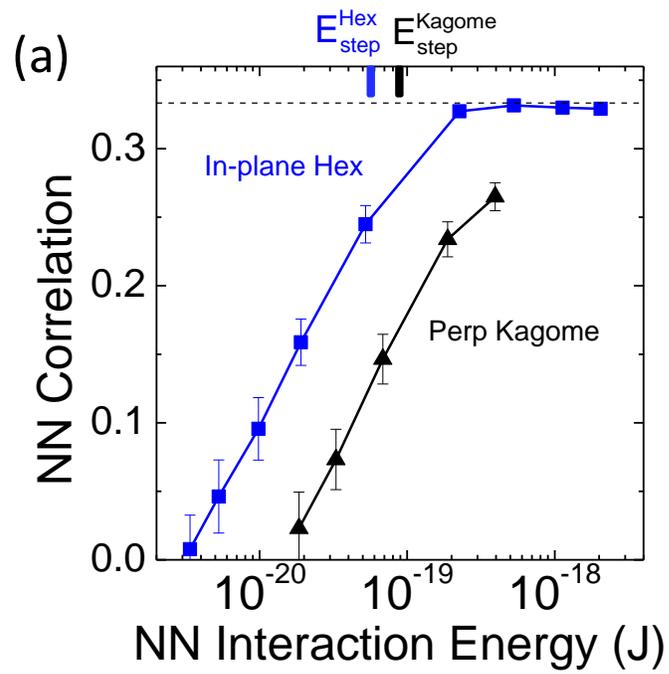

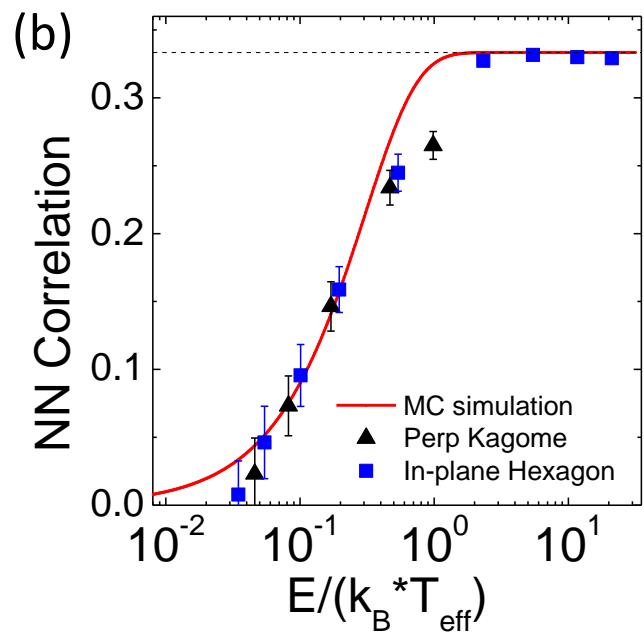

Supplementary Material for: Perpendicular Magnetization and Generic Realization of the Ising Model in Artificial Spin Ice

Sheng Zhang,[1] Jie Li,[1] Ian Gilbert,[1] Jason Bartell,[1] Michael J. Erickson,[2] Yu Pan,[1] Paul E. Lammert,[1] Cristiano Nisoli,[3] K. K. Kohli,[1] Rajiv Misra,[1] Vincent H. Crespi,[1] Nitin Samarth,[1] C. Leighton,[2] and Peter Schiffer[1]

[1]Department of Physics and Materials Research Institute, Pennsylvania State University, University Park, PA 16802, USA

[2]Department of Chemical Engineering and Materials Science, University of Minnesota, Minneapolis, MN 55455, USA

[3]Theoretical Division and Center for Nonlinear Studies, Los Alamos National Laboratory, Los Alamos, NM 87545, USA

**Sample Fabrication and Characterization**

Our samples were fabricated from multilayer stacks with structure Ti(20Å) / Pt($x$) / [Co(3Å)/Pt(10Å)]$_8$ and total thickness of 22.4 nm with $x$ = 100 Å, deposited via electron beam evaporation onto electron-beam patterned bi-layer resist.[1-6] The Co-Co coupling through the Pt interlayer is thought to occur via the RKKY mechanism, with significant similarities to better known cases such as Fe/Cr and Co/Cu GMR systems. The difference in the Co/Pt case is that the Pt is so close to satisfying the Stoner criterion, and is thus so susceptible, that the proximity to the Co induces a small moment on the Pt atoms. This renders the interlayer coupling positive (i.e., ferromagnetic) at all reasonable thicknesses, meaning that the usual oscillatory AF/FM coupling is destabilized. The multilayer films were characterized structurally via Grazing Incidence X-ray Reflectivity (GIXR) and Wide-Angle X-ray Diffraction (WAXRD) in a system equipped with a 2D area detector. GIXR revealed the expected oscillations due to the entire stack thickness in addition to a first order superlattice peak due to the [Co/Pt] superlattice. The actual superlattice period falls in the range 15 to 17 Å, compared to the nominal 13 Å. Intensity versus 2θ scans were obtained by integration of the 2D images in the WAXRD measurement. The only peaks observed were from Si(001), Pt(111), [Co/Pt](111), and associated satellite fringes, confirming (111) texture in the out-of-plane orientation. Fig. S1 shows representative scans for samples with $x$ = 10, 100 and 500 Å, in the vicinity of the Pt buffer layer and [Co/Pt] (111) multilayer reflections. As expected the Pt(111) reflection near 40° decreases with decreasing Pt buffer layer thickness, gradually revealing a shoulder in the vicinity of 40.5°. Consistent with prior literature[1] we ascribe this peak to the [Co/Pt] multilayer, the expected position for [Co(3Å)/Pt(10Å)] being ~40.7°, based on prior work. These data are obtained from radial scans across the 2D images. Tangential scans around the section of the Debye rings visible on the detector (referred to here as χ scans) shown in the inset to the figure (for $x$ = 100 and 500 Å) illustrate the extent of texture. This texturing is significant (though not complete), increasing with increasing Pt buffer layer thickness. This observation provides an explanation for the increasing out-of-plane coercivity with Pt buffer layer thickness, as the effective perpendicular anisotropy in this system is known to improve with (111) texturing[2]. Fig. S2 shows the resulting magnetic anisotropy. We aligned the sample with respect to the field, and we believe that the inflection near zero field is due to a minor component of the total sample volume that has much lower magnetocrystalline anisotropy than the average. This is relatively common in perpendicular magnetic anisotropy materials and can arise due to thickness variations among the various layers, or a progression in microstructure during growth such that either the top or bottom of the multilayer is not optimal. This can even occur in non-multilayered hard magnetic materials, where it is again common (see, for example, Fig. 5 (top panel) of Reference [7]).

**Simulations**

We calculated the interaction energies between islands using the NIST OOMMF (object-oriented micromagnetics framework) code for a pair of islands at different spacings[8, 9]. The OOMMF cell size is 5 × 5 × 5 nm$^3$, comparable to the exchange length of permalloy. The saturation magnetization ($M_S$ = 860 × 10$^3$ A/m) and anisotropy constant ($K1$ = 0) are standard literature values for permalloy. For the [Co(3Å)/Pt(10Å)]$_8$ multilayer, we treated the system as a bulk material with saturation magnetization ($M_S$ = 435 × 10$^3$ A/m) and anisotropy constant ($K1$ = 170 × 10$^3$ J/m$^3$ along the $z$ direction) as extracted from our SQUID measurement, as shown in Fig. S2. Based on the data and measurements of two pieces of the same sample, we estimate the uncertainty at less than 10% in those parameters, which should not qualitatively affect our conclusions.

Simulations to derive the correlations in Fig. 3 of the main text were carried out on systems of 4000 to 8000 spins, which are large enough to eliminate finite-size effects away from criticality. Two models were simulated: A quasi-equilibrium Gibbsian model (Model G) and a kinetic zero-temperature quenched model (Model Z), assuming ideal Ising spins in both cases. Model G was simulated by a standard Metropolis Monte Carlo algorithm. Model Z was simulated by choosing spins at random and flipping them if the nearest-neighbor interaction energy was thereby decreased (and doing nothing if the flip would be energetically neutral). This procedure was continued until the nearest-neighbor correlation reached the desired point. For each stopping point, 10000 to 20000 independent runs were performed. This large a number is not needed for the pair correlations we investigated, but to fill out the tail of the domain size distribution. Note that Model G, being released from the conceptual requirement of equilibrium, can be couched in terms of a more general effective potential function. A nearest-neighbor four-spin term therein can be interpreted kinetically in terms of majority-vote spin flips. Extending the model to include this term improves the model's agreement with the correlations observed in experiment. A more detailed study of the properties of these two models is given elsewhere.[10]

The domains in the honeycomb lattice may reveal subtle long-distance effects that are not evident in the short-range correlations, but should appear in the statistics of the domain sizes. Since we can measure the domain size only within the boundary of a given MFM image of ~1100 islands for 500 nm spacing (or ~260 islands at 1000 nm), we plot in Fig. S3 the fraction of islands within an image that are contained in domains of at least $N$ islands (counting only the visible size for those domains that leak outside the image). The probability of a randomly chosen moment existing within a domain larger than $N$ falls off exponentially with $N$, as expected, at least within the limits of the finite regions imaged by MFM. We gathered the same domain size statistics from the Model G and Model Z simulations, using windows of the same size and shape as the experimental ones. The results shown in Fig. S3, again based on matching only the nearest-neighbor correlations, are in agreement with experiment. This confirms the apparent absence of direct long-range effects and further suggests that the domain structure is insensitive to the precise dynamic mechanisms producing the final array states.

For our comparison of nearest neighbor correlations with the Ising model (Fig. 4b of the main text), we performed Metropolis Monte Carlo simulations for an Ising kagome antiferromagnet with either only nearest-neighbor interaction or dipolar interactions at a series of values of $\beta J$, where $\beta = 1/k_B T_{eff}$, and $J$ is the interaction for nearest neighbors in either case determined by micromagnetic simulations. In each case we ran 15000 samples of a 20×20 lattice, spaced by 12 sweeps through the lattice, after thermalization. There is only a slight difference between the simulations run with dipolar and with nearest-neighbor interactions.

Figure S1. Conventional WAXRD data on Co/Pt multilayers with various Pt buffer layer thickness (10, 100 and 500 Å). The inset shows "tangential" scans around the Debye rings on the 2D area detector.

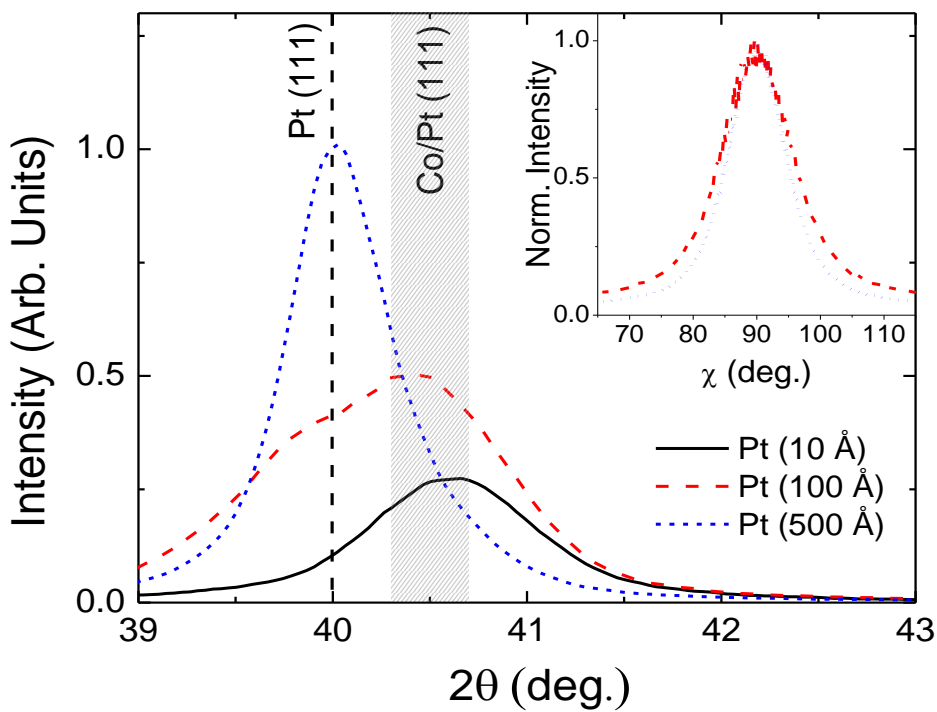

Figure S2. In-plane and out-of-plane hysteresis loops obtained by Superconducting Quantum Interference Device (SQUID) magnetometry on unpatterned Co/Pt multilayer film with 100 Å Pt buffer layer at T = 300 K. The slight curvature in the out-of-plane loop is similar to what would be seen for a slight misalignment of the field to the axis. This effect is, however, most likely due to a minor component of the total sample volume that has much lower magnetocrystalline anisotropy than the average.

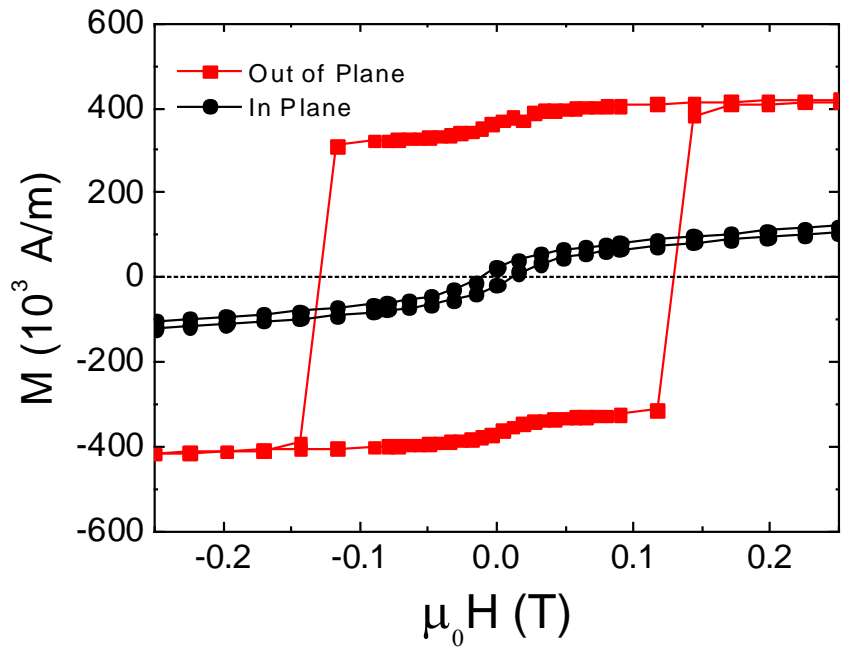

Figure S3. Domain size distribution for different lattice spacings (500nm, 600 nm, 800 nm and 1000 nm) based on five images each. Monte Carlo simulations (windowed to the same size as the MFM scans) in a quasi-equilibrium Gibbsian model (Model G: dense dash lines) and a kinetic zero-temperature quenched model (Model Z: sparse dash lines) both fit the experiments well.

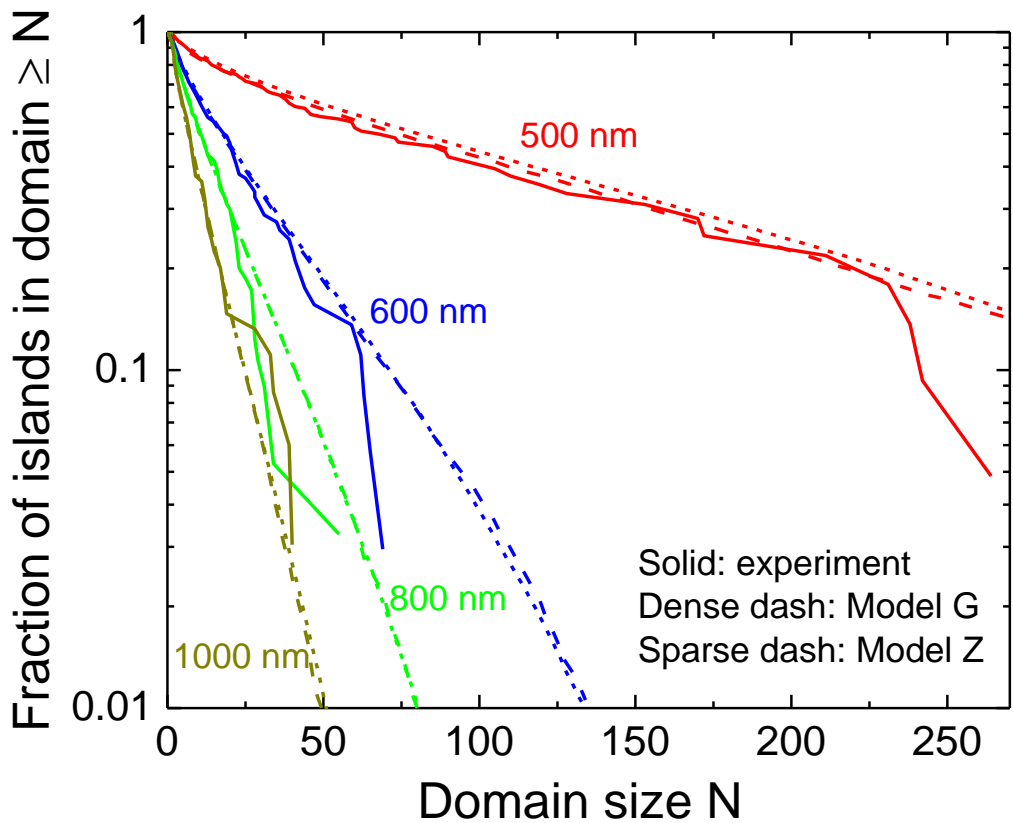